\documentclass[apjl]{emulateapj} 

\def\alwaysmath#1{\ifmmode{#1}\else{$#1$}\fi} 
\newcommand{\kms}{\mbox{km s$^{-1}~$}}
\newcommand{\feh}{[Fe/H]}

\lefthead{Garc\'{\i}a P\'erez et al.}
\righthead{Very Metal-Poor Stars in the Outer Galactic bulge}

\begin{document}

\title{Very Metal-Poor Stars in the Outer Galactic Bulge\\
 Found by the APOGEE Survey}

\author{Ana E. Garc\'{\i}a P\'erez\altaffilmark{1},
Katia Cunha\altaffilmark{2,3},
Matthew Shetrone\altaffilmark{4},
Steven R. Majewski\altaffilmark{1},
Jennifer A. Johnson\altaffilmark{5},
Verne V. Smith\altaffilmark{6},
Ricardo P. Schiavon\altaffilmark{7},
Jon Holtzman\altaffilmark{8},
David Nidever\altaffilmark{9},
Gail Zasowski\altaffilmark{5},
Carlos Allende Prieto\altaffilmark{10,11},
Timothy C. Beers\altaffilmark{6,12},
Dmitry Bizyaev\altaffilmark{13},
Garrett Ebelke\altaffilmark{13},
Daniel J. Eisenstein\altaffilmark{14},
Peter M. Frinchaboy\altaffilmark{15},
L\'eo Girardi\altaffilmark{16,17},
Fred R. Hearty\altaffilmark{1},
Elena Malanushenko\altaffilmark{13},
Viktor Malanushenko\altaffilmark{13},
Szabolcs Meszaros\altaffilmark{11},
Robert W. O'Connell\altaffilmark{1},
Daniel Oravetz\altaffilmark{13},
Kaike Pan\altaffilmark{13},
Annie C. Robin\altaffilmark{18},
Donald P. Schneider\altaffilmark{19},
Mathias Schultheis\altaffilmark{18},
Michael F. Skrutskie\altaffilmark{1}, 
Audrey Simmonsand\altaffilmark{13}, and
John C. Wilson\altaffilmark{1}}

\altaffiltext{1}{Department of Astronomy, University of Virginia, Charlottesville, VA 22904-4325, USA}
\altaffiltext{2}{Steward Observatory, University of Arizona, Tucson, AZ 85721, USA} 
\altaffiltext{3}{Observat\'orio Nacional, S\~ao Crist\'ov\~ao, Rio de Janeiro, Brazil}
\altaffiltext{4}{University of Texas at Austin, McDonald Observatory, Fort Davis, TX 79734, USA}
\altaffiltext{5}{Department of Astronomy, The Ohio State University, Columbus, OH 43210, USA}
\altaffiltext{6}{National Optical Astronomy Observatories, Tucson, AZ 85719, USA}
\altaffiltext{7}{Gemini Observatory, 670 N. A'Ohoku Place, Hilo, HI 96720, USA}
\altaffiltext{8}{New Mexico State University, Las Cruces, NM 88003, USA}
\altaffiltext{9}{Department of Astronomy, University of Michigan, Ann Arbor, MI 48109, USA} 
\altaffiltext{10}{Departamento de Astrof\'{\i}sica, Universidad de La Laguna, 38206 La Laguna, Tenerife, Spain}
\altaffiltext{11}{Instituto de Astrof\'{\i}sica de Canarias, E38205 La Laguna, Tenerife, Spain}
\altaffiltext{12}{Department of Physics \& Astronomy and JINA, Joint Institute for Nuclear Astrophysics, Michigan State University, E. Lansing, MI 48824, USA}
\altaffiltext{13}{Apache Point Observatory, PO Box 59, Sunspot, NM 88349-0059, USA}
\altaffiltext{14}{Harvard Smithsonian Center for Astrophysics, 60 Garden Street, Cambridge, MA 02138, USA}
\altaffiltext{15}{Texas Christian University, Fort Worth, TX 76129, USA}
\altaffiltext{16}{Laborat\'orio Interinstitucional de e-Astronomia - LIneA, Rua Gal. Jos\'e Cristino 77, Rio de Janeiro, RJ - 20921-400, Brazil}
\altaffiltext{17}{Osservatorio Astronomico di Padova, INAF, Vicolo dell'Osservatorio 5, I-35122 Padova, Italy}
\altaffiltext{18}{Institut Utinam, CNRS UMR6213, OSU THETA, Universit\'e de Franche-Comt\'e, 41bis avenue de lÕObservatoire, 25000 Besan\c{c}on, France} 
\altaffiltext{19}{Department of Astronomy and Astrophysics, The Pennsylvania State University, University Park, PA 16802}

\begin{abstract}

Despite its importance for understanding the nature of early
stellar generations and for constraining Galactic bulge formation models, 
at present little is known about the metal-poor
stellar content of the central Milky Way. This is a consequence of
the great distances involved and intervening dust obscuration,
which challenge optical studies. However, the Apache Point Observatory
Galactic Evolution Experiment (APOGEE), a wide-area, multifiber,
high-resolution spectroscopic survey within Sloan Digital Sky Survey
III (SDSS-III), is exploring the chemistry of all Galactic stellar
populations at infrared wavelengths, with particular emphasis on
the disk and the bulge. An automated spectral analysis of data on
2,403 giant stars in twelve fields in the bulge obtained during
APOGEE commissioning yielded five stars with low metallicity
([Fe/H]$\le-1.7$), including two that are very metal-poor [Fe/H]$\sim-2.1$
by bulge standards.

Luminosity-based distance estimates place the five stars
within the outer bulge, where other 1,246 of the analyzed stars 
may reside. A manual reanalysis of the spectra
verifies the low metallicities, and finds these stars to be enhanced
in the $\alpha$-elements O, Mg, and Si without significant
$\alpha$-pattern differences with other local halo or metal-weak
thick-disk stars of similar metallicity, or even with other more
metal-rich bulge stars. While neither the kinematics nor chemistry
of these stars can yet definitively determine which, if any, are
truly bulge members, rather than denizens of other populations
co-located with the bulge, the newly-identified stars reveal that
the chemistry of metal-poor stars in the central Galaxy resembles
that of metal-weak thick-disk stars at similar metallicity.

\end{abstract}


\section{Introduction}

The chemical compositions of the oldest stars hold fundamental clues about the 
early history of galaxies. Even if no true Population III stars presently exist in the Milky Way, 
their nature can be constrained from observations of the 
elemental abundance patterns of the most metal-poor existing 
Galactic stars \citep[e.g.,][]{Beers05,Ekstrom2008}. 
Theoretical predictions suggest that the oldest,
most metal-poor stars in the Milky Way (MW) 
are to be found in the bulge \citep[e.g.,][]{White00, Tumlinson10}.
However, these ancient relics are extremely difficult to identify
because of the combination of very high extinction, foreground contamination, 
and the fact that the most metal-poor stars are but a small fraction of 
a large population of stars located in the inner Galaxy.
As a result, our view of the early stages of Galactic formation and chemical 
evolution has been skewed by studies of more easily accessible 
Galactic halo samples, at large Galactocentric distances.

To date, the origin of the Galactic bulge is still uncertain. 
The boxy X-shape \citep[e.g.,][]{McWilliam2010} and
high metallicity \citep[e.g.,][]{Rich88} of the bulge suggests
secular formation associated with the disk and bar --- i.e., a {\it
pseudo}-bulge \citep[e.g.,][]{Immeli04, Kormendy04}. On the other
hand, the metallicity gradient of the bulge seen by \citet{Zoccali08}
and its old age \citep[$\sim$10~Gyr, ][]{Clarkson08} are more 
consistent with a {\it classical} bulge
\citep[e.g.,][]{Rahimi10,Bournaud11}. However, the above criteria
are clearly not definitive model discriminators, given that:
\citet{Kunder12} also discuss metallicity gradients in a secular
scenario, the high metallicity in the classical scenario could be
explained by early starbursts \citep{McWilliam94}, and old age could
be understood in a secular scenario if disk instabilities occurred
at early times. Moreover, recent studies by \citet{Babusiaux10} 
and \citet{Hill11} suggest that the central MW may include the
superposition of a classical bulge {\it and} a pseudo-bulge.

To discriminate between these formation scenarios, a comprehensive 
chemical analysis of the central MW is needed, to characterize the
metal-poor end of its metallicity distribution function (MDF) and
assess abundance patterns of bulge populations at all metallicities.
The ratio between $\alpha$-element and iron abundances 
([$\alpha$/Fe]\footnote{[X/Fe]=$A(X)-A(X)_{\sun}-(A(\mathrm{Fe})-A(\mathrm{Fe})_{\sun})$,
$A(X)=\log({N_X}/{N_{\mathrm{H}}})+12$, where $N_{X}$ represents
the number density of nuclei of element X.}) of a stellar population is sensitive 
to the initial mass function of its parental
population, whereas the position of the ``knee'' of the metallicity-[$\alpha$/Fe]
relation is sensitive to the early star formation rate \citep[e.g.,][]{McWilliam97}. 
The {\it spread} in [$\alpha$/Fe] at given [Fe/H] depends on whether 
the metal-poor bulge stars were accreted or produced {\it in situ},
and by which mechanisms \citep{Immeli04,Rahimi10}. Moreover, in
the secular instability scenario, the bulge [$\alpha$/Fe] pattern
should resemble that of the inner disk. Unfortunately, this
discriminatory power of chemical abundances has barely been exploited
because most spectroscopic studies have been restricted to high
metallicity bulge stars. For example, the pioneering medium-resolution
optical study of twelve giants by \citet{McWilliam94} that discovered
the bulge to be $\alpha$-enhanced (a signature of rapid formation)
was limited to \feh$\ge-1.08$. Similar results were obtained from
high-resolution, near-infrared spectroscopy of fourteen bulge giants
with \feh$\sim-0.33$ by \citet{Rich05}, and seven more with
\feh$\ge-1.05$ by \citet{Cunha06}. Subsequent high-resolution
optical analyses (e.g., dozens of stars by \citealt{Fulbright07},
\citealt{Zoccali06}, and \citealt{Lecurer07}) --- still probing
only \feh$\ge-1.30$ --- revealed a bulge that is more 
$\alpha$-enhanced than the local thick disk. Starbursts were invoked to explain these higher $\alpha$-element
levels within a classical formation scenario. 
However, more recent studies comparing bulge with inner disk 
\citep{Bensby10}, or thick-disk stars 
\citep[and with more homogeneous analyses ---][]{Melendez08,Ryde10,Alves10,Gonzalez11} 
{\it did} find common 
abundance patterns, which supports bulge formation scenarios invoking either
secular evolution or radial stellar migrations associated with 
spiral arms and/or the bar \citep[][but cf. 
\citealt{Minchev12}]{Schoenrich09}.

A striking aspect of all previous spectroscopic surveys of the bulge
is that despite sample sizes approaching a thousand stars, 
until only very recently the most metal-poor star identified 
had \feh$= -1.69$ \citep{Zoccali08}, with only four stars having 
\feh$<-1.5$ known. Clearly, any hope of probing the extremely minor,
but exceedingly interesting, metal-poor content of the central
Galaxy requires much larger samples --- a challenging prospect,
given the significant distance and foreground dust obscuration. 
The situation is changing rapidly, the large ARGOS survey at
medium-resolution recently reported {\feh} and averaged [$\alpha$/Fe]
for stars in the inner Galaxy down to \feh$=-2.60$ \citep{Ness12}.
Here we report the discovery of five additional stars with 
[Fe/H]$\lesssim-1.5$ in the central Galaxy found within a sample of $\sim$2,403
stars observed in bulge fields by the Apache Point Observatory Galactic Evolution Experiment (APOGEE; \citealt{Majewski2010}),
part of the Sloan Digital Sky Survey III \citep[SDSS-III;][]{Eisenstein11},
commissioning. APOGEE uses a high resolution, $H$-band spectrograph with 300 optical fibers mated to 
the large field-of-view, Sloan 2.5-m telescope \citep{Gun06}.
We also present detailed abundance ratios for these stars and find
that they are similar to metal-poor stars in other parts of the
Galaxy. 

\placetable{log}

\section{Observations and Abundance Analysis}

\begin{deluxetable*}{lrrrrrr}
\tablecolumns{6}
\setlength{\tabcolsep}{0.06in}
\tabletypesize{\scriptsize}
\tablecaption{Derived Stellar Parameters and Abundances for the Metal-Poor Bulge Candidates.}
\tablenum{1}
\tablehead{\colhead{2MASS Star ID = }  & \colhead{17062946-2325097} &  \colhead{17083699-2257328} & \colhead{17324728-1735240} &  \colhead{18013387-1907266} &  \colhead{18155672-2133077}}
\startdata
$l$ [\degr] &  359.727099 & 0.396365  &  8.108990  & 10.301410    &  9.811497  \\
$b$ [\degr]   &  10.358167 & 10.228863  &     8.491406  &  1.842886     &   $ -2.282258$ \\
$\alpha_{2000}$ [h m s]  &17 06 29.46 &17 08 36.99 & 17  32 47.28  & 18 01 33.87  & 18 15 56.72\\
$\delta_{2000}$ [h m s] &$-23$ 25 09.7 &  $-22$ 57 32.8 &   $-17$ 35 24.0 &  $-19$  07 26.6  & $-21$ 33 07.7\\
 $H$ [mag] &  9.38 & 9.794  &  9.686 &  9.665 & 8.828 \\
$A_{\rm Ks}$ [mag] &  0.319 & 0.293 &  0.247 &  0.607 &  0.275 \\
$V_{helio}$ [\kms]  & $-39.49$ & $328.49$ &  $21.03$ & $142.17$ & $-49.79$\\
$d$ [kpc]    &    9.43 &  8.37  &     9.64  &     7.40  &     5.71   \\
$S/N$  & 403 & 326  & 230  & 159  &251  \\
$T_{eff}$ [K]  & 3900 ($\pm150$) & 4300 ($\pm150$)  & 4200 ($\pm150$) & 4000 ($\pm150$)  & 4100 ($\pm150$) \\
$\log{g}$ [cgs]  & 0.36 ($\pm0.50$) & 0.70 ($\pm0.50$)  & 0.55 ($\pm0.50$)  & 0.52 ($\pm0.50$) & 0.63 ($\pm0.50$)\\
$[{\rm Fe/H}]$ & $-1.47\ (\pm 0.20)$ & $-2.10\ (\pm 0.20) $  & $-2.05\ (\pm 0.20)$  & $-1.54\ (\pm 0.20)$  & $-1.66\ (\pm 0.20)$\\
$\xi_t$ [{\kms}]  & 3.0 ($\pm0.5$) & 2.5 ($\pm0.5$)  & 2.5 ($\pm0.5$)  & 2.5 ($\pm0.5$)  & 2.5 ($\pm0.5$)\\
$A ({\rm Fe})$ &$5.98\ (\pm 0.12)$ & $5.35\ (\pm 0.12)$ & $5.40\ (\pm 0.12)$ & $5.91\ (\pm 0.12)$  &$5.79\ (\pm 0.12)$\\
$A ({\rm O})$ & $7.72\ (\pm 0.38)$ &$7.04\ (\pm 0.38)$  & $7.13\ (\pm 0.38)$  & $7.75\ (\pm 0.38)$  & $7.42\ (\pm 0.38)$ \\
$A ({\rm Mg})$ & $6.23\ (\pm 0.15)$& $5.76\ (\pm 0.15)$ & $5.71\ (\pm 0.15)$ & $6.30\ (\pm 0.15)$ & $6.12\ (\pm 0.15)$ \\
$A ({\rm Si})$ & $6.15\ (\pm 0.10)$& $5.58\ (\pm 0.10)$ &  $5.57\ (\pm 0.10)$ &  $6.36\ (\pm 0.10)$  & $6.18\ (\pm 0.10)$ \\
$[\rm{O/Fe}]$  & $+0.53\ (^{+0.28}_{-0.26})$ & $+0.48\ (^{+0.28}_{-0.26}) $  & $+0.52\ (^{+0.28}_{-0.26})$  & $+0.63\ (^{+0.28}_{-0.26})$  & $+0.42\ (^{+0.28}_{-0.26})$\\
$[\rm{Mg/Fe}]$ & $+0.17\ (^{+0.09}_{-0.07})$ & $+0.33\ (^{+0.09}_{-0.07})$  & $+0.23\ (^{+0.09}_{-0.07})$  & $+0.31\ (^{+0.09}_{-0.07})$  & $+0.25\ (^{+0.09}_{-0.07})$\\
$[\rm{Si/Fe}]$ & $+0.11\ (^{+0.05}_{-0.08})$ & $+0.17\ (^{+0.05}_{-0.08})$  & $+0.11\ (^{+0.05}_{-0.08}) $  & $+0.39\ (^{+0.05}_{-0.08})$  & $+0.33\ (^{+0.05}_{-0.08}) $\\
\enddata
\label{stepar}
\end{deluxetable*}

\begin{figure}
\figurenum{1}
\plotone{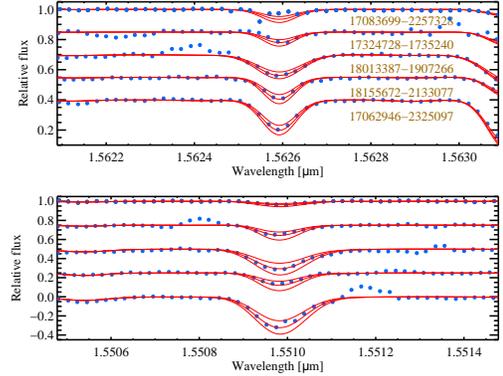}
\caption{Observed (circles) and synthetic (solid line) spectra of the Fe~{\sc i}~1.56259~{$\mu$}m (top panel) and OH lines at $\sim$1.5510~{$\mu$}m (bottom panel) for the five metal-poor stars in our study. Spectra were offset vertically by multiples of 0.15 (Fe) and 0.25 (OH) for clarity. The synthesis for the best-fitting abundance and $\pm$0.2~dex from that are also shown.}
\label{fig1}
\end{figure}

APOGEE commissioning observations were taken in May-July 2011 for
$\sim$4,700 K/M giant stars in 18 fields spanning
$-1${\degr}$<l<20${\degr}, $|b|<20${\degr}, and $\delta>-32${\degr}
(see \citealt{Nidever12}, Fig.\ 1). Stars were selected from the
2MASS Point Source Catalog \citep{Skrutskie06} by color ([$J-K_{\rm
s}]_0\ge0.5$) and magnitude ($H$$\leq$11.0) (see Zasowski et al.
2013). The observed spectra were of high quality ($R=22,500$,
$S/N>150$ per pixel, at near Nyquist sampling), although misplacement 
of the red detector led to degraded resolution
($R\sim$14,500) for $1.65<\lambda<1.70\mu$m. The raw datacubes were reduced to calibrated, 
1-D spectra and stellar radial velocities (RVs) were derived using the
APOGEE reduction pipeline \citep{Nidever12}. Effective temperatures 
($T_{\rm{eff}}$), surface gravities ($\log{g}$), and [Fe/H] from an early
version of the APOGEE Stellar Parameter and Chemical Abundances Pipeline 
(ASPCAP, Garc{\'{\i}}a P{\'e}rez et al.2013, in prep.) 
were used to select candidate metal-poor stars in 12 
bulge commissioning fields within 10.5{\degr} from the
Galactic center. Six stars were selected as having
$\mathrm{[Fe/H]}_{\rm{ASPCAP}}\le-1.7$, but one was rejected from
further consideration for showing peculiar line profiles (potentially
a spectroscopic binary). Specific sections of the APOGEE spectra
of the five metal-poor stars (Table~\ref{stepar}) were then re-analyzed
interactively via a classical 1D-LTE spectrum synthesis approach (see Fig.~\ref{fig1}). 
The synthesis used {\sc marcs} model atmospheres \citep{Asplund97},
computed for the individual stellar parameters and chemical
compositions listed in Table~\ref{stepar}. Equipped with the model
atmospheres, stellar spectra were synthesized using the Uppsala
code {\sc bsyn} and a line list (version 201202161204) compiled
specifically for APOGEE (Shetrone et al. 2013, in prep.). Both the 
instrumental and macroturbulence profiles were described by Gaussians
whose widths were adjusted to the variable instrumental resolution
($\lambda/\Delta\lambda=12,000$--$22,000$). Following 
\citet[][]{Melendez08}, several iterations were performed to ensure
consistency between the derived chemical compositions and those of the model
atmospheres employed.

Initial estimates of the atmospheric parameters were based on the
observed spectra, in combination with theoretical isochrones. 
To determine $T_{\rm{eff}}$, the ratio ($R_{\mathrm{OH-Mg}}$) of
the sum of OH line strengths (at 1.57589~$\mu$m and 1.57608~$\mu$m)
to that of a nearby Mg~{\sc i} line (at 1.57533~$\mu$m) was
measured.\footnote{Cited wavelengths refer to vacuum measurements.}
$R_{\mathrm{OH-Mg}}$ is quite sensitive to $T_{\rm{eff}}$, due to
the strong temperature dependence of OH for $T_{\rm{eff}}\gtrsim$
4500~K, with only a small dependence on $\log{g}$. We calibrated
$R_{\mathrm{OH-Mg}}$ versus $T_{\rm{eff}}$ using data for the field
red giants $\alpha$ Boo, $\mu$ Leo, $\beta$ And, and $\delta$ Oph,
and giants from the globular clusters M3, M13, and M71. Though
spanning a large range in metallicity, age, and mass, these
particular stars define an $R_{\mathrm{OH-Mg}}$-$T_{\rm{eff}}$
relation with an intrinsic scatter of only $\sim\pm$100--120~K.

Surface gravities were derived from isochrones \citep{Dartmouth08}
with an assumed age of 10~Gyr and [$\alpha$/Fe]=$+0.6$ (consistent
with our final derived values). The adopted values of $T_{\rm{eff}}$
and $\log{g}$ for the stars are given in Table~\ref{stepar}, and
were checked using the temperature and gravity-sensitive profiles
of H~{\sc i} lines at 1.61137~{$\mu$}m and 1.68111~{$\mu$}m, with
theoretical line absorption profiles from \citet{Ali66}.

Stellar metallicity estimates ([Fe/H]) are based on mean values 
of iron abundance derived from a sample of four to thirteen measured 
Fe~{\sc i} lines, and assuming a solar abundance value of
$A$(Fe)$_\sun$=7.45 \citep{Asplund05}. For the other elements, 
values of $A$(O)$_\sun$=8.66, $A$(Mg)$_\sun$=7.58, and
$A$(Si)$_\sun$=7.55 \citep{Asplund05} were assumed. The lines were
selected from the ASPCAP line list among those with minimum
blending from molecular lines in the atmospheric parameter range 
explored in this study. Sample spectra and synthesis
for the Fe~{\sc i} ~1.56259~{$\mu$}m line are shown in Figure~\ref{fig1}.

Microturbulent velocities ($\xi_t$) were derived by forcing 
consistency between the abundances obtained from weak and strong
Mg and Si lines. The lines used were the following: Mg~{\sc
i} 1.57450, 1.57533, 1.57700, and 1.59588~{$\mu$}m, and Si~{\sc
i} 1.59644, 1.60992, 1.66853, and 1.66853~{$\mu$}m. We obtain
$\xi_t = 2.5$~km s$^{-1}$ for all stars (except one with 3.0 km
s$^{-1}$).

The oxygen abundances were obtained from the analysis of 10 to 17
OH lines covering $\lambda\lambda=1.5395$--1.6376~{$\mu$}m. The
mean abundance values are listed in Table~\ref{stepar}, 
and the observed and synthetic spectra of OH lines at
$\sim$1.5510~{$\mu$}m are shown in Fig.~\ref{fig1} for all stars.
Because there may be some interdependence of O and C abundances
through CO formation, our determinations require {\it a priori}
knowledge of C abundances, which were estimated from very weak
CO bands, and the non-detection of the C~{\sc i} atomic line at
1.68950~{$\mu$}m in any of the stars.


Internal errors in the abundances were derived from the
abundance sensitivity to stellar parameters (Table~\ref{sensit}) 
using the star 2M17083699-2257328 as a baseline and adopting the
values listed in Table~2 as our uncertainties in the other parameters.
For all elements, and oxygen in particular, abundance uncertainties 
are most sensitive to errors in $T_{\rm{eff}}$. Overall, the
abundances show moderate sensitivity to errors in $\log{g}$ 
(typically $\Delta$ $A(X)<0.1$~dex), and are similarly or less 
sensitive to uncertainties in microturbulence and [Fe/H]. Final
uncertainties were computed by adding the errors in quadrature and
are 0.12, 0.38, 0.15, and 0.10~dex for Fe, O, Mg, and Si, respectively.

\begin{table}
\tablenum{2}
\begin{center}
\caption{Abundance Sensitivity to Stellar Parameter Uncertainties. \label{sensit}}
\begin{tabular}{rrrrrrrrrr}
& $A ({\rm Fe})$ & $A ({\rm O})$ & $A ({\rm Mg})$ &$A ({\rm Si})$ \\
\tableline

$\Delta T_{eff} (+150~K)$  & 0.090 & 0.276 & 0.090 & 0.072\\
                               ($-150~K$) &$-0.084$ &$-0.258$&$-0.090$&$-0.042$\\
$\Delta \log{g}$ (0.5) [cgs] &$-0.008$ &$-0.110$&$-0.060$&0.000\\
                                      ($-0.5$) [cgs] &0.032&0.140&0.050&0.015\\
$\Delta \xi_t$  (0.5~\kms)  &$-0.008$&$-0.006$&$-0.080$&$-0.050$\\
                         ($-0.5$~\kms)&0.016&0.006&0.080&0.070\\
$\Delta [{\rm Fe/H}]$ (0.2~dex) &0.003&0.120&$-0.007$ &0.006\\
                             ($-0.2$~dex) &0.007&$-0.100$&0.005&0.000\\
\end{tabular}
\end{center}

\label{log}
\end{table}


\section{Population Membership}

The stars in Table~1 have ($l$, $b$) typical of the outer bulge, as do 
$\sim1246$ other automatically-analyzed stars in our sample, but it is unclear whether 
the stars in that table are actually in, and belong to, the bulge. To gauge distances, 
luminosities were estimated from the adopted
$\log{g}$ and derived $T_{\rm{eff}}$, assuming $M$=0.8~$M_{\sun}$,
as expected for old giants. To determine $M_H$, bolometric corrections
were estimated from $T_{\rm{eff}}$ using a calibration derived
from stellar isochrones in \citet{Girardi00}. Extinctions were estimated by
combining near- and mid-IR photometry \citep{Majewski11} and the
\citet{Indebetouw05} extinction law. The derived distances
(Table~\ref{stepar}) have relatively large uncertainties, given the
uncertainties in $T_{\rm{eff}}$, gravities, assumed masses, and extinctions.
\begin{figure}
\figurenum{2}
\plotone{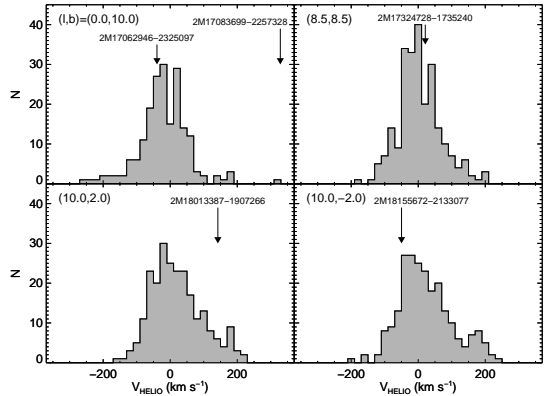}
\figcaption[fig2.eps]{Heliocentric RV distributions for stars in
the four observed bulge fields, with the velocities of the five
metal-poor stars indicated.}
\label{fig2}
\end{figure}
Nevertheless, the distances --- projected on the Galactic plane at
$d/{\rm cos}(b)$=5.71--9.59~kpc --- place these stars marginally
or completely within the nominal bulge, assuming the latter has a
$\sim$2--3~kpc radius and $\sim$8~kpc distance. However, both the
Besan\c{c}on \citep{Robin12} and Trilegal \citep{Vanhollebeke09} 
MW models predict that APOGEE target selection in these ``bulge"
fields should also yield a small number of metal-poor {\it halo} stars
--- though the expected ratio of metal-poor halo to bulge stars is
presently unconstrained because it is highly dependent on uncertain
extrapolations of the halo density law to small Galactocentric
radii, as well as on the unknown shape of the bulge MDF. Moreover,
RVs provide little additional discrimination between bulge and 
overlapping halo populations because of the similar (near zero)
mean velocity and comparably large velocity dispersions of the two
populations. The measured RVs are generally {\it compatible} with
those of more metal-rich bulge stars dominating the samples in these
fields (Fig.~{\ref{fig2}), although the star 2M17083699-2257328 has
an extreme velocity ($V_{\rm{hel}}=328.5$~\kms) compared to bulge
stars in the same field ($<V_{\rm{hel}}>=-18.1$~\kms, $\sigma_v$ =
53.4~\kms) and may therefore less likely be a bulge star on dynamical 
grounds. We conclude that, while our metal-poor stars are
spatially coincident with the bulge-dominated, central Galaxy, we
cannot definitively ascribe population membership to them by position
or velocity.

\section{The Iron and $\alpha$-Element Content of the Metal-Poor Stars}

Even if as many as four of the five stars in this study are 
truly bulge members, they would represent a mere 0.17\% of the 2,403
candidate red-giant stars targeted in the twelve bulge fields.
Careful, detailed, analysis of the data confirmed their
low metallicities, with three at [Fe/H]$\sim-1.6$, and two at
[Fe/H]$\sim-2.1$}. The metallicities derived here are robust 
([Fe/H] errors $\pm0.11$~dex), and comparable to, or 
lower than, the median metallicities of either local halo or metal-weak
thick-disk stars, but certainly much more metal poor than the typical
bulge star. Whether they are bulge, halo, or even thick-disk
members, these stars are among the most metal poor ever found in 
the central parts of the Galaxy.

\begin{figure}
\figurenum{3} \plotone{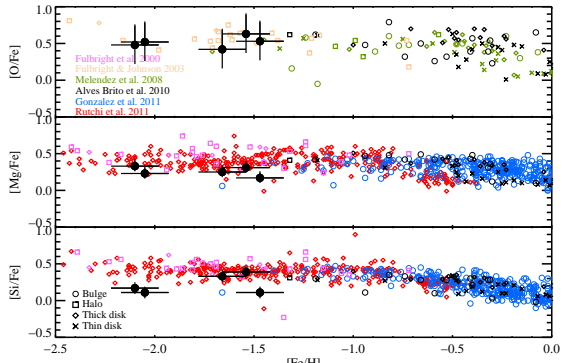} \figcaption[fig3.ps]{Comparison of
our $\alpha$-element measurements of the Table~1 stars (filled
circles) with those from the literature for bulge, disk, and halo
stars.} \label{fig3}
 \end{figure}

Our abundance estimates suggest that all five stars are $\alpha$-enhanced,
with mean abundance ratios and standard deviations [O/Fe]=$+0.52\pm0.08$,
[Mg/Fe]=$+0.26\pm0.06$, and [Si/Fe] =$+0.22\pm0.13$. Figure~\ref{fig3}
contains these derived [X/Fe] (along with literature values ---
rescaled to our assumed solar abundances --- for bulge, disk, and
halo stars), and shows Si to have the most scatter (with perhaps a
hint of two [Si/Fe] subgroups), but oxygen to be most enhanced.
A range of solar oxygen values exists in the literature
(from different indicators and/or modeling); a different choice
would have led to smaller or even larger enhancements. 
The $A$(O) are also the most uncertain because of the great sensitivity
of molecular line formation to atmospheric structure and, therefore,
to the modeling employed in the spectral synthesis and to the adopted stellar parameters (especially 
$T_{\rm{eff}}$). For abundance {\it ratios}, part of the sensitivity
to stellar parameters is cancelled out, so that the [O/Fe], [Mg/Fe],
and [Si/Fe] internal errors shown in Figure~\ref{fig3} are reduced
to $^{+0.28}_{-0.26}$, $^{+0.09}_{-0.07}$ and $^{+0.05}_{-0.08}$~dex,
respectively.

With the exception of silicon, the $\alpha$-enhancement abundances 
of our sample stars are not much different from those of other,
previously reported ``metal-poor bulge stars" shown in Figure~\ref{fig3},
although the latter are almost entirely at higher metallicity and
exhibit significant scatter. Three of our stars exhibit lower Si
enhancements, but comparable to those seen in the only available
literature datapoint for [Fe/H]$<-1.5$ \citep{Gonzalez11}. However, the
latter star also apparently has a peculiar, smaller Mg enhancement,
a feature shared with none of our stars. Data for more metal-poor
stars in the Galactic bulge are needed to confirm whether these
particular enhancement variations are a distinctive feature of
low-metallicity stars in the bulge.

Assuming our sample stars are true bulge members, it is interesting 
to compare their abundances with those of thick-disk stars.
\citet{Fulbright07} and \citet{Zoccali06}, deriving large
$\alpha$-enhancements for the bulge compared to published thick-disk
abundances, argue for a higher star formation rate for the bulge,
whereas the studies by \citet{Melendez08}, \citet{Alves10}, and
\citet{Gonzalez11} (which analyzed stars from both populations
homogeneously) claim no significant $\alpha$-enhancement differences
(\S1). We can now extend these comparisons to lower metallicities
using the disk data from \citet{Fulbright00},
\citet{Fulbright03}\footnote{Oxygen abundances based only on the
forbidden [O~{\sc i}] line at 630~nm were used from this source.},
and \citet{Ruchti11}. Bear in mind that such a comparison, especially
for oxygen (see \citealt{Garcia06}), should be viewed with caution
because of potential systematic errors in the analyses: different
abundance and stellar parameter scales, different stellar evolutionary
stages, different abundance indicators, and different locations in
the Galaxy. With these caveats, we find the abundances of our stars
to be comparable to those of the metal-weak thick-disk stars in
Figure~\ref{fig3}. While (1) some of our Si abundances may be
marginally too low, and (2) there are not many metal-weak thick-disk
points for comparison to our oxygen results, the metallicities and
$\alpha$-element abundance patterns of our low-metallicity stars
are comparable to what is found in the metal-weak thick disk. This
makes a somewhat stronger case for a possible connection between
the bulge and thick disk, as suggested previously, and lends further
support to the notion that migrating stars populated both the
Galactic bulge and thick disk \citep[][]{Schoenrich09}.

The results in Table~1 indicate that these stars, as a group, do
not have unusual [O/Fe] values compared to those of halo stars, but
may contain some members that have lower [Si/Fe] and slightly lower
[Mg/Fe] values. Indeed, three of the stars have low values of
[Si/Fe] compared to most halo stars of similar metallicity, with
one being 2M17083699-2257328, which has the most extreme RV and
thus might be expected to be the most likely halo member. The two
remaining stars cannot be chemically distinguished from local halo
stars. It should be noted that \citet{Nissen10} identified a
population of $\alpha$-poor halo stars. More data on the metal-poor
populations of the inner Galaxy may help to disentangle possible
metal-poor bulge stars from halo stars.

\section{Conclusions}

We have found five giant stars within the commissioning data of the
SDSS-III's APOGEE project that have sky positions and Galactic
plane-projected distances ($d/{\rm cos}(b)$=5.71--9.59~kpc) expected
for the bulge, but that exhibit distinctly low iron content ($-2.10\le
\mathrm{[Fe/H]}\le-1.47$). We present abundance ratios for
these stars, significantly augmenting the sample of metal-poor bulge
stars with detailed chemical information and including two stars
much more metal-poor ([Fe/H]$\sim-2.1$) than the previous bulge
star with this information \citep[a micro-lensed dwarf with
\feh$=-1.89$,][]{Bensby12}, which was excluded from our comparison
of only giants. 

There is no strong evidence that our stars are significantly
chemically different from other more metal-rich bulge stars --- 
or even different from other stars in the halo or metal-poor thick
disk, although some stars in our sample do exhibit somewhat lower
Si enhancements than typically seen in other Galactic stars at
these metallicities. Unfortunately the presently available kinematics
and chemistry are not sufficient to determine with certainty how
many of the stars may be true bulge members. Nevertheless, this
initial APOGEE sample significantly contributes to the task of
compiling a more throrough census of the metal-poor stellar content
of the central MW, and portends the promising results to be expected
from the ongoing APOGEE exploration of the Galactic bulge.

\acknowledgments

We acknowledge funding from NSF grants AST11-09718 and AST-907873. 
Funding for SDSS-III has been provided by the Alfred P. Sloan
Foundation, the Participating Institutions, the National Science
Foundation, and the U.S. Department of Energy Office of Science. The
SDSS-III website is http://www.sdss3.org/.
SDSS-III is managed by the Astrophysical Research Consortium for the
Participating Institutions of the SDSS-III Collaboration including the
University of Arizona, the Brazilian Participation Group, Brookhaven
National Laboratory, University of Cambridge, Carnegie Mellon
University, University of Florida, the French Participation Group, the
German Participation Group, Harvard University, the Instituto de
Astrofisica de Canarias, the Michigan State/Notre Dame/JINA
Participation Group, Johns Hopkins University, Lawrence Berkeley
National Laboratory, Max Planck Institute for Astrophysics, Max Planck
Institute for Extraterrestrial Physics, New Mexico State University,
New York University, Ohio State University, Pennsylvania State
University, University of Portsmouth, Princeton University, the
Spanish Participation Group, University of Tokyo, University of Utah,
Vanderbilt University, University of Virginia, University of
Washington, and Yale University.

\clearpage

\end{document}